\documentclass[Times,8pt,aps,prl,onecolumn,showpacs,amsmath,amssymb,floatfix,footinbib,superscriptaddress]{revtex4-1}
\usepackage{amsmath,natbib,graphicx,subfigure,amssymb,graphics,amsmath,mathrsfs,CJK,color}
\usepackage{multirow,fancyhdr,color,bm,tabularx,psfrag,geometry,dcolumn,datetime}
\usepackage[colorlinks=true,linkcolor=blue,urlcolor=blue,citecolor=blue]{hyperref}
\usepackage[mathlines]{lineno}
\def\footnoterule{\kern -1mm \hrule width 5.8cm \kern 2.2mm}%
\usepackage{tikz,xcolor,hyperref}
\definecolor{lime}{HTML}{A6CE39}
\DeclareRobustCommand{\orcidicon}{%
 \begin{tikzpicture}
 \draw[lime, fill=lime] (0,0)
    circle [radius=0.16]
    node[white] {{\fontfamily{qag}\selectfont \tiny ID}};\draw[white, fill=white] (-0.0625,0.095)
    circle [radius=0.007];
 \end{tikzpicture}
\hspace{-2mm}}
\foreach \x in {A, ..., Z}
{\expandafter\xdef\csname orcid\x\endcsname{\noexpand\href{https://orcid.org/\csname orcidauthor\x\endcsname}{\noexpand\orcidicon}}}

\begin{document}
\title{Zero absorption and large negative refractive index in a left-handed four-level atomic media}

\author{ShunCai Zhao\textsuperscript{\orcidA{}}}
\email[Corresponding author]{zscnum1@126.com}
\affiliation{School of Materials Science and Engineering, Nanchang University, Nanchang 330031,PR China}
\affiliation{Engineering Research Center for Nanotechnology, Nanchang University, Nanchang 330047,PR China}
\affiliation{Institute of Modern Physics,Nanchang University,Nanchang 330031,PR China}
\author{ZhengDong Liu}
\email[Corresponding author]{lzdgroup@ncu.edu.cn}
\affiliation{School of Materials Science and Engineering, Nanchang University, Nanchang 330031,PR China}
\affiliation{Engineering Research Center for Nanotechnology, Nanchang University, Nanchang 330047,PR China}
\affiliation{Institute of Modern Physics,Nanchang University,Nanchang 330031,PR China}

\author{Qixuan Wu}
\affiliation{English department,Hainan University,Danzhou 571737,PR China}
\begin{abstract}
In this paper,we have investigated three external fields interacting with the four-level atomic system described by the density-matrix
approach.The atomic system exhibits left-handedness with zero absorption as well as large negative refractive index.Varying the
parameters of the three external fields,the properties of zero absorption,large negative refractive index from the atomic system
keep unvarying.Our scheme proposes an approach to obtain negative refractive medium with zero absorption. The zero absorption property
of atomic system may be used to amplify the evanescent waves that have been lost in the imaging by traditional lenses.And a slab
fabricated by the left-handed atomic system may be an ideal candidate for designing perfect lenses.
\end{abstract}
\keywords{zero absorption; left-handedness; large negative refractive index; negative permittivity; negative permeability}

\maketitle
\section{Introduction}
The propagation of electromagnetic waves in matter is characterized by the frequency-dependent relative dielectric permittivity
$\varepsilon_{r}$ and magnetic permeability $\mu_{r}$. And their product defines the index of refraction:$\varepsilon_{r}$$\cdot$
$\mu_{r}$$=$$n^{2}$. The left-handed material(LHM) has a negative refractive index with the permittivity and permeability being
negative simultaneously[1]. And there're much remarkable progress demonstrates negative refractive index using technologies such as
artificial composite metamaterials [2-14], transmission line simulation[15], photonic crystal structures[16-21] and photonic
resonant materials [22-26]. The former three methods, based on the classical electromagnetic theory, require delicate manufacturing of
spatially periodic structure. The last method is a quantum optical approach where the physical mechanism is the quantum interference
and coherence that arises from the transition process in a multilevel atomic system. The LHM becomes a very active area of
research because of one of the goals of ''perfect lens" in which imaging resolution is not limited by the wavelength [27]. However, the
results in Ref.[28-29]show that the LHM lenses can indeed amplify the evanescent waves that have been lost in the imaging by
traditional lenses,the presence of absorption(even a small amount) plays a crucial rule in recovery of the lost evanescent
waves,thereby controlling the quality of imaging,and makes LHM lenses less perfect. So, the key challenge remains the realization of
negative refraction without absorption, which is particularly important in the optical regime[30-32].

With the realization of negative refractive index without absorption in mind, here we propose a promising new approach: the use of quantum
interference effects causing electromagnetically induced transparency(EIT)[33]to realize left-handedness with zero
absorption. Under some appropriate conditions,the system shows negative values for permittivity and permeability simultaneously
, zero absorption as well as a large negative refractive index.Our approach for left-handedness without absorption is different from
Ref.[34]and our scheme allows us to engineer the value of the refractive index while maintaining vanishing absorption[35] to the
beam.In this sense, our approach implements modifying the optical response of an atomic medium.

The paper is organized as follows.In Section 2, we present our model and its expressions for the electric permittivity,magnetic
permeability and refractive index.In Section 3, we present numerical results and their discussion. This is followed by concluding remarks
in Section 4.

\section{Theoretical model}

\begin{center}
\begin{figure}[h!]
  \centering
  \includegraphics[width=3.0in]{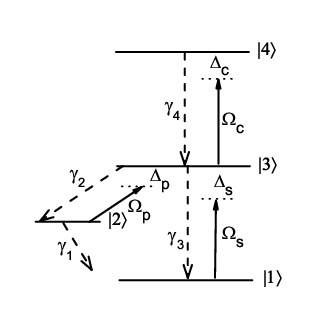}%
  \hspace{0in}%
  \caption{Schematic diagram of a four-level atomic system
interacting with three electromagnetic
fields,$\Omega_{c}$,$\Omega_{p}$ and $\Omega_{s}$. The level pairs
$|2\rangle$-$|3\rangle$ and $|2\rangle$-$|1\rangle$ are coupled to
the electric and magnetic fields of $\Omega_{p}$ (the weak probe
light), respectively.}
\end{figure}\label{Fig.1}
\end{center}

Consider a four-level atomic ensemble interacting with three optical fields, i.e. the coupling beam,probe light and signal field. The
configuration of such a four-level system is depicted in Fig.1. The coupling field $E_{c}$ (frequency $\omega_{c}$) is acting on the
transition $|3\rangle$ and $|4\rangle$ (with transition frequency $\omega_{34}$).The weak probe field $E_{p}$(frequency
$\omega_{p}$)interacts with the transition $|2\rangle$ to $|3\rangle$ (transition frequency $\omega_{23}$),and its electric
and magnetic fields are coupled to the level pairs
$|3\rangle$-$|2\rangle$ and $|2\rangle$-$|1\rangle$, respectively. While the weak signal field $E_{s}$(frequency
$\omega_{s}$)drives the transition $|1\rangle$ to $|3\rangle$(transition frequency $\omega_{13}$). The frequency detunings of these
three optical fields are $\Delta_{c}$,$\Delta_{p}$ and $\Delta_{s}$, respectively. The Rabi frequencies of these optical fields are
denoted by $\Omega_{c}$,$\Omega_{p}$ and $\Omega_{s}$, respectively. All the Rabi frequencies are assumed to be real. We consider the two
levels, $|3>$ and $|2>$ have opposite parity with $d_{23}=<2|\vec{\hat{d}}|3>\neq0$ ,and $\vec{\hat{d}}$ is the electric dipole operator. The levels $|2\rangle$ and $|1\rangle$
have the same parity, $\mu_{12}=<1|\vec{\hat{\mu}}|2>\neq0$, where the $\vec{\hat{\mu}}$ is the magnetic dipole operator. The radiative
decay constants from levels $|4\rangle$ to $|3\rangle$, $|3\rangle$ to $|2\rangle$ and $|3\rangle$ to $|1\rangle$ are
$\gamma_{4}$, $\gamma_{2}$, $\gamma_{3}$. $\gamma_{1}$ is related to non-radiative relaxation of state $|2\rangle$.Under the condition
that lower levels are equally populated and the decay rate of level $|1\rangle$ is quite small, then levels $|1\rangle$,$|3\rangle$ and
$|4\rangle$ are in a three-level ladder-type configuration and level $|2\rangle$, together with levels $|3\rangle$ and $|4\rangle$ forms
another three-level ladder-type configuration.

For this system the Hamiltonian under the dipole and rotating-wave
approximation (RWA) can be written as,
\begin{equation}
\hat{H}=\Delta_{p}|2\rangle\langle2|+
\Delta_{s}|1\rangle\langle1|+\Delta_{c}|4\rangle\langle4|+[\Omega_{p}|3\rangle\langle2|+\Omega_{s}|3\rangle\langle1|+\Omega_{c}|3\rangle\langle4|+H.c.]
\end{equation}
Using the density-matrix approach, the time-evolution of the system
is described as
\begin{equation}
\frac{d\rho}{dt}=-\frac{i}{\hbar}[H,\rho]+\Lambda\rho ,\label{eq1}
\end{equation}
In which $\Lambda\rho$ represents the irreversible decay part in the
system which is a phenomenological added decay term that corresponds
to the incoherent processes.And the density-matrix equations of
motion in the dipole and RWA for this system can be written as
follows,

\begin{equation}
\dot{\rho_{11}}=2\gamma_{3}\rho_{33}+2\gamma_{1}\rho_{22}+i\Omega{s}(\rho_{13}-\rho_{31})
\end{equation}
\begin{equation}
\dot{\rho_{22}}=2\gamma_{2}\rho_{33}-2\gamma_{1}\rho_{22}+i\Omega{p}(\rho_{23}-\rho_{32})
\end{equation}
\begin{equation}
\dot{\rho_{33}}=-2(\gamma_{2}+\gamma_{3})\rho_{33}+2\gamma_{4}\rho_{44}-i\Omega{s}(\rho_{13}-\rho_{31})-i\Omega{p}(\rho_{23}-\rho_{32})+i\Omega{c}(\rho_{34}-\rho_{43})
\end{equation}
\begin{equation}
\dot{\rho_{44}}=-2\gamma_{4}\rho_{44}-i\Omega{c}(\rho_{34}-\rho_{43})
\end{equation}
\begin{equation}
\dot{\rho_{12}}=-[\gamma_{1}-i(\Delta_{s}-\Delta_{p})]\rho_{12}+i\Omega{p}\rho_{13}-i\Omega{s}\rho_{32}
\end{equation}
\begin{equation}
\dot{\rho_{13}}=-(\gamma_{2}+\gamma_{3}-i\Delta_{s})\rho_{13}+i\Omega{p}\rho_{12}+i\Omega{c}\rho_{14}+i\Omega{s}(\rho_{11}-\rho_{33})
\end{equation}
\begin{equation}
\dot{\rho_{14}}=-[\gamma_{4}-i(\Delta_{s}+\Delta_{c})]\rho_{14}+i\Omega{c}\rho_{13}-i\Omega{s}\rho_{34}
\end{equation}
\begin{equation}
\dot{\rho_{23}}=-(\gamma_{2}+\gamma_{3}-i\Delta_{p})\rho_{23}+i\Omega{s}\rho_{21}+i\Omega{c}\rho_{24}+i\Omega{p}(\rho_{22}-\rho_{33})
\end{equation}
\begin{equation}
\dot{\rho_{24}}=-[\gamma_{1}+\gamma_{4}-i(\Delta_{p}+\Delta_{c})]\rho_{24}+i\Omega{c}\rho_{23}-i\Omega{p}\rho_{34}
\end{equation}
\begin{equation}
\dot{\rho_{34}}=-(\gamma_{2}+\gamma_{3}+\gamma_{4}-i\Delta_{c})\rho_{34}-i\Omega{s}\rho_{14}-i\Omega{p}\rho_{24}+i\Omega{c}(\rho_{33}-\rho_{44})
\end{equation}

Where the frequency detunings of these three optical fields are
$\Delta_{c}=\omega_{c}-\omega_{34}$,
$\Delta_{p}=\omega_{p}-\omega_{23}$ and
$\Delta_{s}=\omega_{s}-\omega_{13}$, respectively.

In the following, we will discuss the electric and magnetic responses of the medium to the probe field.It should be noted that
here the atoms are assumed to be nearly stationary(e.g., at a low temperature) and hence any Doppler shift is neglected. When discussing
how the detailed properties of the atomic transitions between the levels are related to the electric and magnetic susceptibilities, one
must make a distinction between macroscopic fields and the microscopic local fields acting upon the atoms in the vapor. In a
dilute vapor, there is little difference between the macroscopic fields and the local fields that act on any atoms(molecules or group
of molecules)[36]. But in dense media with closely packed atoms(molecules), the polarization of neighboring atoms(molecules)
gives rise to an internal field at any given atom in addition to the average macroscopic field, so that the total fields at the atom are
different from the macroscopic fields[36].In order to achieve the negative permittivity and permeability,here the chosen vapor with
atomic concentration $N=5\times10^{24}m^{-3}$ should be dense, so that one should consider the local field effect, which results from
the dipole-dipole interaction between neighboring atoms. In what follows we first obtain the atomic electric and magnetic
polarization, and then consider the local field correction to the electric and magnetic susceptibilities(and hence to the permittivity
and permeability)of the coherent vapor medium. With the formula of the atomic electric polarizations
$\gamma_{e}=2d_{23}\rho_{23}/\epsilon_{0}E_{p}$, where $E_{p}=\hbar\Omega_{p}/d_{23}$ one can arrive at

\begin{eqnarray}
\gamma_{e}=\frac{2d_{23}^2\rho_{32}}{\epsilon_{0}\hbar\Omega_{p}}\
\end{eqnarray}

In the similar fashion, by using the formulae of the atomic magnetic polarizations $\gamma_{m}=2\mu_{0}\mu_{12}\rho_{21}/B_{p}$ [36], and
the relation of between the microscopic local electric and magnetic fields $E_{p}/B_{p}=c$ we can obtain the explicit expression for the
atomic magnetic polarizability. Where $\mu_{0}$is the permeability of vacuum, c is the speed of light in vacuum.Then,we have obtained the
microscopic physical quantities $\gamma_{e}$and$\gamma_{m}$. In order to achieve a significant magnetic response, the transition frequency
between levels $|2\rangle$-$|3\rangle$, and $|2\rangle$-$|1\rangle$ should be approximately equal to the frequency of the probe
light. Thus, the coherence $\rho_{12}$ drives a magnetic dipole, while the coherence $\rho_{23}$ drives an electric dipole. However, what we
are interested in is the macroscopic physical quantities such as the electric and magnetic susceptibilities which are the electric
permittivity and magnetic permeability. The electric and magnetic Clausius-Mossotti relations can reveal the connection between the
macroscopic and microscopic quantities.According to the Clausius-Mossotti relation [36], one can obtain the electric
susceptibility of the atomic vapor medium

\begin{eqnarray}
\chi_{e}=N\gamma_{e}\cdot{{{{(1-\frac{N\gamma_{e}}{3})}}}}^{-1}
\end{eqnarray}

The relative electric permittivity of the atomic medium reads $\varepsilon_{r}=1+\chi_{e}$. In the meanwhile, the magnetic
Clausius-Mossotti [37]

\begin{eqnarray}
\gamma_{m}=\frac{1}{N}(\frac{\mu_{r}-1}{\frac{2}{3}+\frac{\mu_{r}}{3}})
\end{eqnarray}

shows the connection between the macroscopic magnetic permeability $\mu_{r}$ and the microscopic magnetic polarizations $\gamma_{m}$. It
follows that the relative magnetic permeability of the atomic vapor medium is

\begin{eqnarray}
\mu_{r}=\frac{1+\frac{2}{3}N\gamma_{m}}{1-\frac{1}{3}N\gamma_{m}}
\end{eqnarray}

Substituting the expressions of $\varepsilon_{r}$ and $\mu_{r}$ into $n=-\sqrt{\varepsilon_{r}\mu_{r}}$ [1], we can get the refractive
index of left-handed materials. In the above, we obtained the expressions for the electric permittivity, magnetic permeability and
refractive index of the four-level atomic medium.In the section that follows, we will get solutions to the density-matrix equations
(3)-(11) under the stead-state condition.

\section{Results and discussion}

\begin{center}
\begin{figure}[h!]
  \centering
  \includegraphics[width=3.1in]{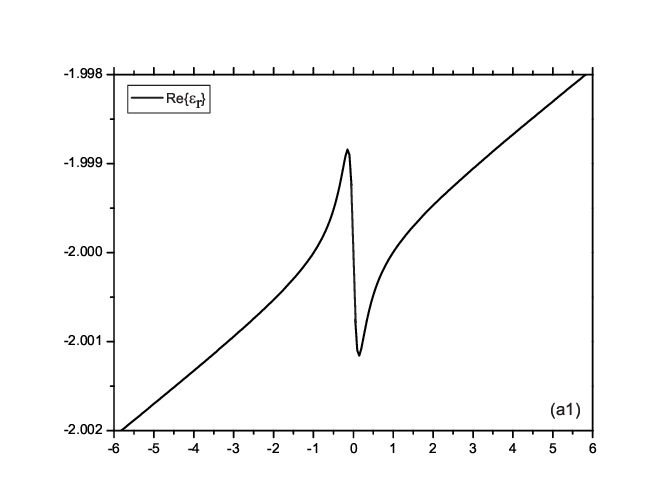}%
  \hspace{0in}%
  \includegraphics[width=3.0in]{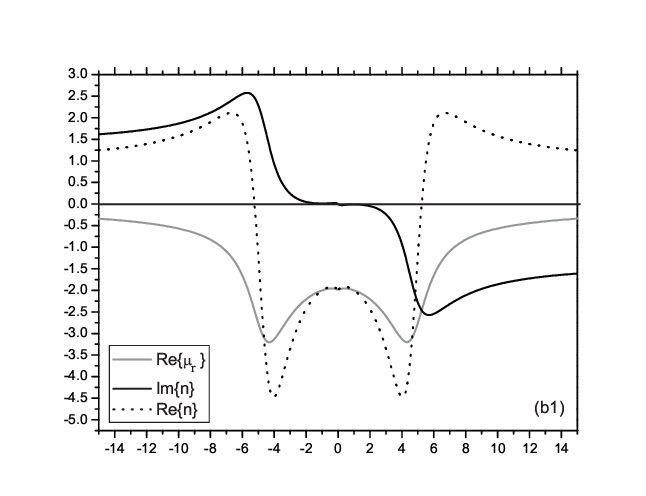}
  \hspace{0in}%
  \includegraphics[width=3.12in]{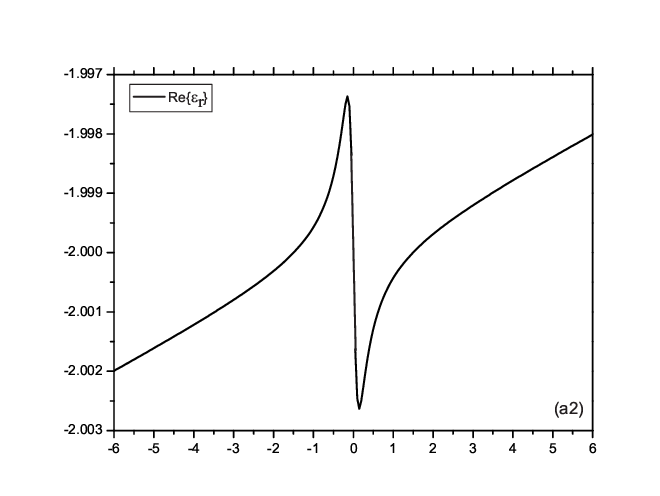}
  \hspace{0in}%
  \includegraphics[width=3.0in]{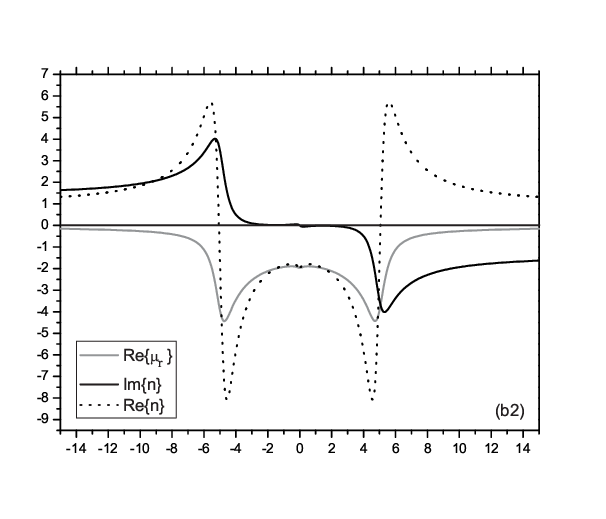}
  \hspace{0in}%
  \includegraphics[width=3.1in]{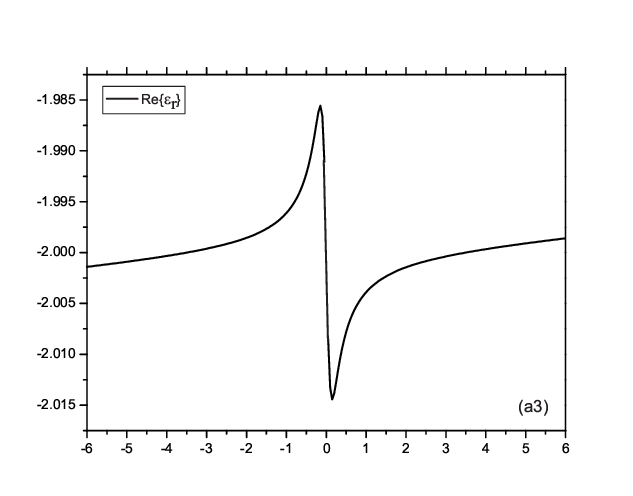}
  \hspace{0in}%
  \includegraphics[width=3.0in]{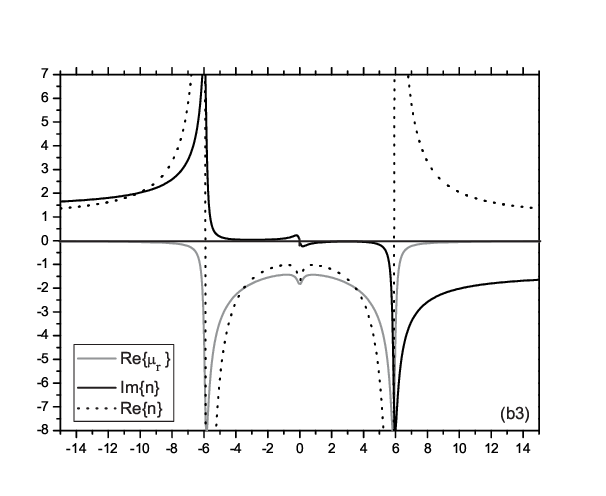}
  \hspace{0in}%
  \includegraphics[width=3.1in]{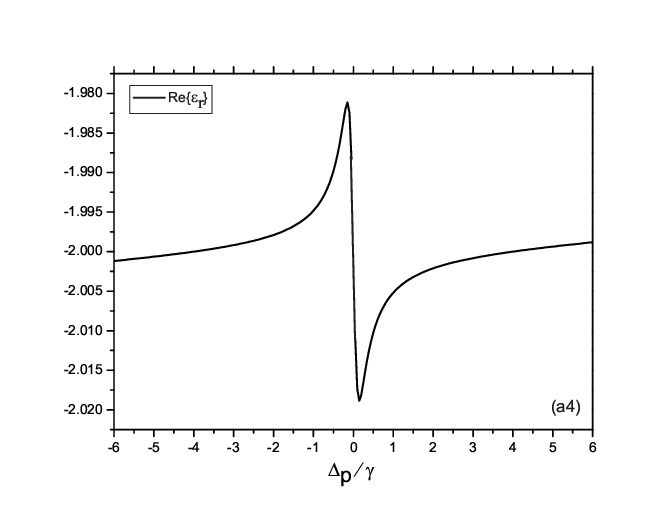}
  \hspace{0in}%
  \includegraphics[width=3.0in]{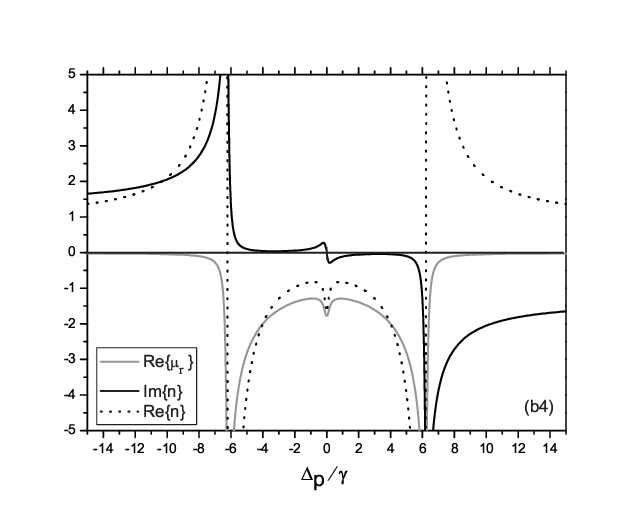}
  \hspace{0in}%
  \caption{Real parts of the permittivity and permeability, the refractive
index as a function of $\Delta_{p}$/$\gamma$ with different
$\Omega_{c}$=$1.0\gamma$, $1.5\gamma$, $3.5\gamma$, $4.0\gamma$.Other
parameters are: $\Delta_{c}$ =$\Delta_{s}$=$0$, $\gamma_{1} $=
$0.05\gamma$, $\gamma_{2}$=$\gamma_{3}$=$0.01\gamma$, $\gamma_{4}$=$0.1\gamma$,
  $\Omega_{p}$=$\Omega_{s}$=$0.1\gamma$.}
\end{figure}\label{Fig.2}
\end{center}

We firstly consider the coupling laser beam and signal laser beam are assumed to be on resonance simultaneously, and the atomic vapor
medium with $N=5\times10^{24}m^{-3}$ to examine the case of dense gas for the four level atom configuration. As pointed out in the
following numerical example, the magnitude of the applied coherent optical field's Rabi frequency is $(10^{8}/s)$,which is larger than
the radiative decay constants. Thus, the effects of dephasing, and collisional broadening can be neglected, and the coherence effect
can be maintained even increased in the atomic vapor under consideration. With the steady solution of the matrix
equations(3)-(11) we can obtain the coherent terms $\rho_{32}$ and $\rho_{21}$,and with the expressions for the atomic electric and
magnetic polarizability (13)-(16) we will present a numerical example to show that the strong electric and magnetic responses can truly
arise in the four-level coherent vapor medium under certain conditions. The strong electric and magnetic responses can lead to
simultaneously negative permittivity and permeability at a certain wide frequency bands of the probe light. The real parts of the
relative permittivity and permeability, refractive index are plotted in Figure 2, Figure 3 and Figure 4. We examine the left-handedness of
the four-level atomic vapor. Largely, we concerned about the absorption behaviors and the values of the negative refractive index
in the zero absorption intervals.

In Figure 2, we set the coupling field and signal field being resonant simultaneously, i.e., $\Delta_{c}$=$\Delta_{s}$=$0$. Other
parameters are scaled by$\gamma$=$10^{8}/s$: $\gamma_{1}$=$0.05\gamma$, $\gamma_{2}$=$\gamma_{3}$=0.01$\gamma$, $\gamma_{4}$=$0.1\gamma$,
$\Omega_{p}$= $\Omega_{s}$ =$0.1\gamma$.Re$[\varepsilon_{r}]$ is
plotted in Figure 2(a1) $\rightarrow$(a4), $Re[\mu_{r}]$ and the
refractive index n are plotted in Figure2(b1)$\rightarrow$(b4)versus
the probe detuning $\Delta_{p}$/$\gamma$ with the Rabi frequency
$\Omega_{c}$= $1.0\gamma$, $1.5\gamma$, $3.5\gamma$,$4.0\gamma$. And
the solid, gray and dotted curves represent $Im[n]$, Re[$\mu_{r}$]and
$Re[n]$in Figure2(b1)$\rightarrow$(b4), respectively. It can be seen
from Figure 2(a1)$\rightarrow$(a4)that the real parts of the
relative dielectric permittivity keep negative values unchanged
while varying the couple Rabi frequency. And the profiles are
symmetry at the exact resonant location.The values of
Re[$\mu_{r}$](gray curves in Figure 2(b1)$\rightarrow$(b4))are
negative too. Thus, the atomic system displays left-handedness with
simultaneous negative permittivity and permeability. The absorption
behaviors of the four-level atomic system are depicted by the
imaginary part(the solid curve in Figure 2(b1)$\rightarrow$ (b4))of
refractive index n. From the profile of the solid curves, we can see
the zero absorption intervals widening with the variation of the
couple field.And the intervals are about [-2$\gamma$, 2$\gamma$](in
Figure 2(b1)), [-3$\gamma$,3$\gamma$](in Figure 2(b2)). When the Rabi
frequency $\Omega_{c}$=$3.5\gamma$, the zero absorption interval is
split into two:[-4.6$\gamma$, -1$\gamma$] and
[1$\gamma$,4.6$\gamma$](in Fig.2(b3)), approximately. In Figure
2(b4), the zero absorption intervals are
[-4.6$\gamma$, -1.8$\gamma$]and [1.8$\gamma$, 4.6$\gamma$], which are
narrow compared with those in Figure 2(b3). The dotted curves
indicate that the real part of refractive index have negative
values. And the maximums of $Re[n]$$\approx$-2.5(in(b1)),
$\approx$-3(in (b2)), $\approx$-4(in (b3))and $\approx$-2.8(in
(b4))in the zero absorption intervals. The results indicate that the
negative refractive index increase with the intensities of the
control field, but high intensity may reduce the values while keeping
zero absorption property unchanged. In the proof-of-principle
experiment to enhance imaging resolution made by N.A.Proite et
al.[35], the maximum refractive index of an atomic vapor can reach to
2 approximately while keeping without absorption.Here we achieve
zero absorption in the negative refractive index media, which maybe a
possibility to overcome the defect mentioned in Ref.[28-29] and to
enhance the resolution of image.

\begin{center}
\begin{figure}[h!]
  \hspace{0in}%
  \centering
 \includegraphics[width=3.1in]{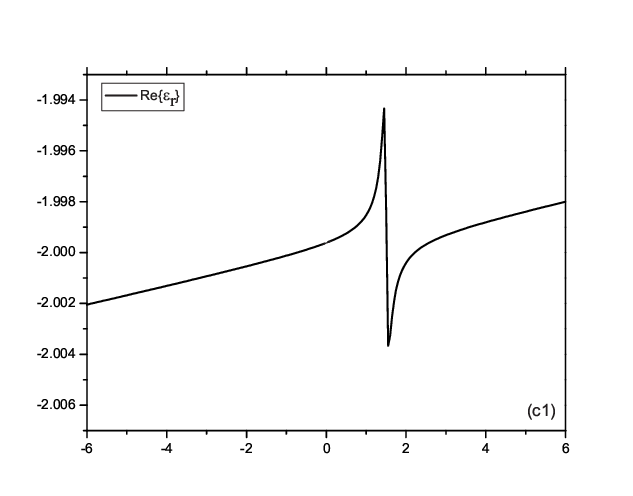}
  \hspace{0in}%
  \includegraphics[width=3.0in]{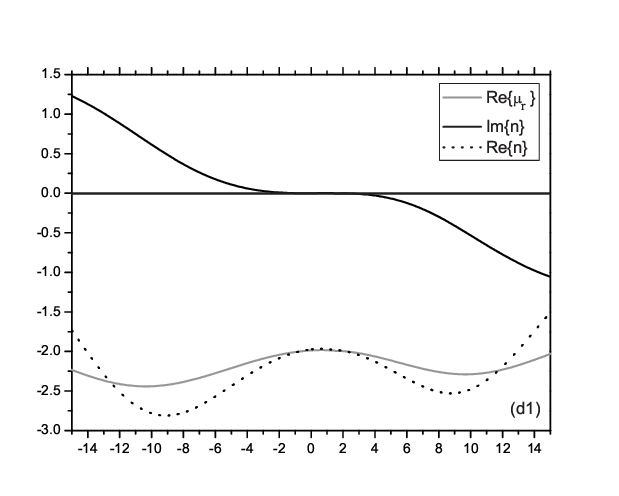}
  \hspace{0in}%
\includegraphics[width=3.1in]{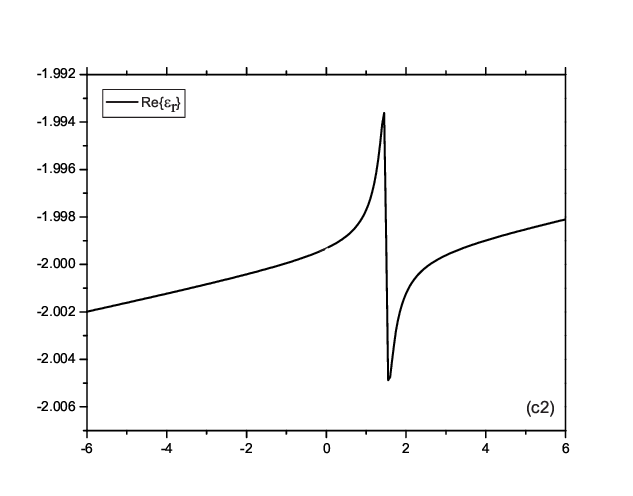}
  \hspace{0in}%
  \includegraphics[width=3.0in]{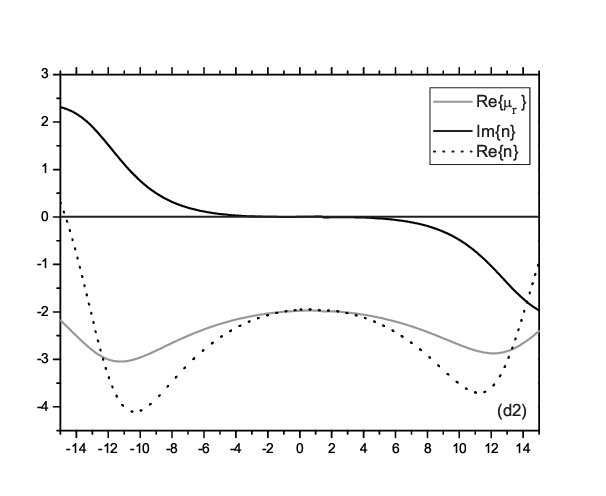}
  \hspace{0in}%
\includegraphics[width=3.1in]{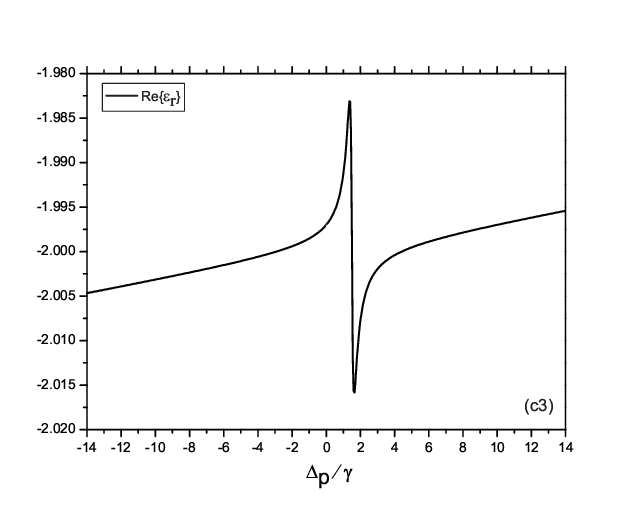}
  \hspace{0in}%
  \includegraphics[width=3.0in]{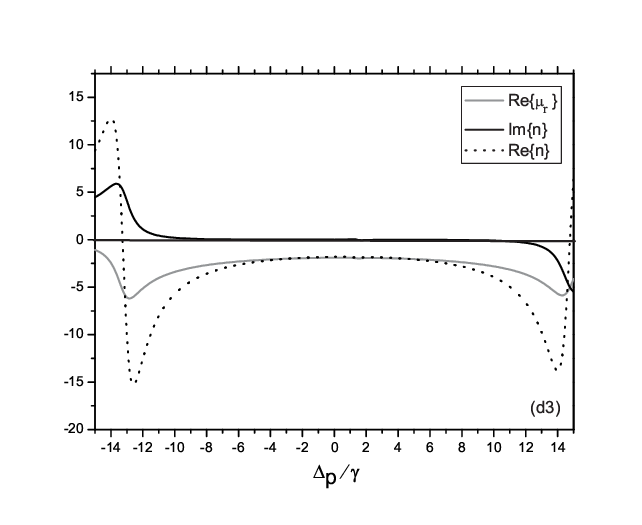}
  \hspace{0in}%
  \caption{Real parts of the permittivity and permeability,the refractive
  index as a function of $\Delta_{p}$/$\gamma$ with: $\Delta_{c}$=$-1.5\gamma$, $\Delta_{s}$=$1.5\gamma$,
  $\Omega_{p}$=$\Omega_{s}$=$0.8\gamma$, $\Omega_{c}$=$1.0\gamma$, $1.5\gamma$,
  $3.5\gamma$. Other parameters are the same as those in Figure 2.}
\hspace{0in}%
\end{figure}\label{Fig.3}
\end{center}

Figure 3 gives the absorption behaviors and the left-handedness when
the coupling and signal fields are simultaneous
off-resonant, i.e., the frequency detunings $\Delta_{c}$=-1.5$\gamma$,
$\Delta_{s}$ =1.5$\gamma$.In Figure3(c1)
$\rightarrow$(c3), $Re[\varepsilon_{r}]$ is plotted corresponding to
the different Rabi frequencies($\Omega_{c}$):$1.0\gamma$,
$1.5\gamma$,$3.5\gamma$. We can see the negative values of
$Re[\varepsilon_{r}]$ and the unsymmetrical character of the
profiles.T he values of the real part of relative permeability
depicted by gray curves are negative in Figure 3(b1)$\rightarrow$
(b4). The atomic system still displays left-handedness when the
signal and coupling fields are simultaneous off-resonant. We notice
that the values of the imaginary part of the refractive
index(depicted by the solid curves in Figure 3(d1)$\rightarrow$
(d3))are zero in the intervals of [-2.5$\gamma$, 3.0$\gamma$](in
(d1)),[-3.0$\gamma$, 3.5$\gamma$](in (d2))and [-9.0$\gamma$,
11.2$\gamma$](in (d3)). The zero absorption intervals are wider than
those under the condition of the coupling field and signal field
being resonant simultaneously in Figure 2. And the couple field is
stronger,the zero absorption intervals are wider. The wide zero
absorption intervals may provide the possibility to facilitate
making "perfect" images. And the maximum values of negative
refractive index are about -2.2 in (d1), -2.5 in (d2), -6 in (d3)in
the zero absorption intervals.An effective positive refractive index
of the designing three-dimensional isotropic
metamaterials[38]reaches between 5.5 and 7 made by J. W. Shin et
al.. Here we get a counter-intuitive negative values of the
refractive index with zero absorption and the maximum negative value
is -6, approximately.

\begin{center}
\begin{figure}[h!]
  \centering
  \includegraphics[width=3.1in]{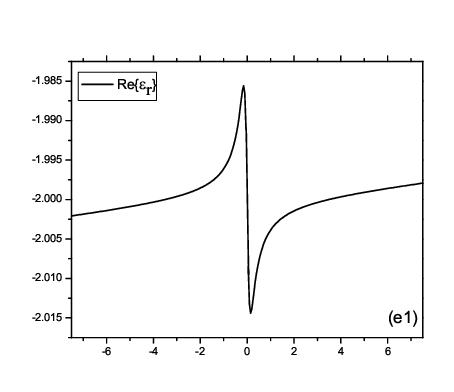}
  \hspace{0in}%
  \includegraphics[width=3.0in]{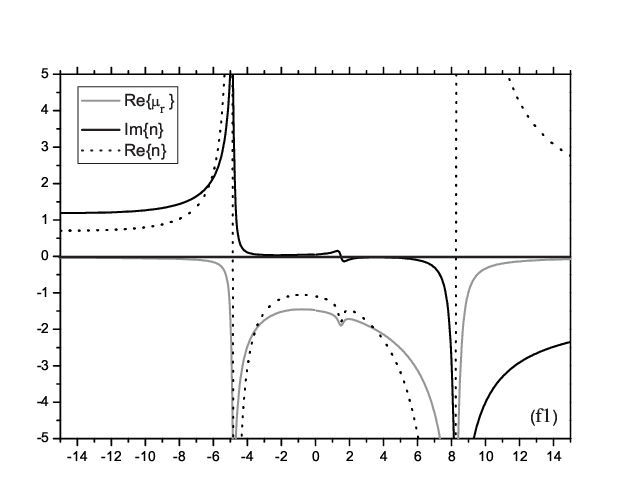}
  \hspace{0in}%
  \includegraphics[width=3.1in]{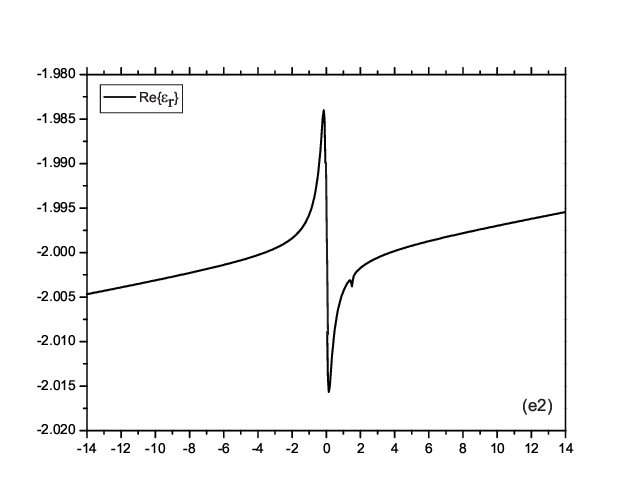}
  \hspace{0in}%
  \includegraphics[width=3.0in]{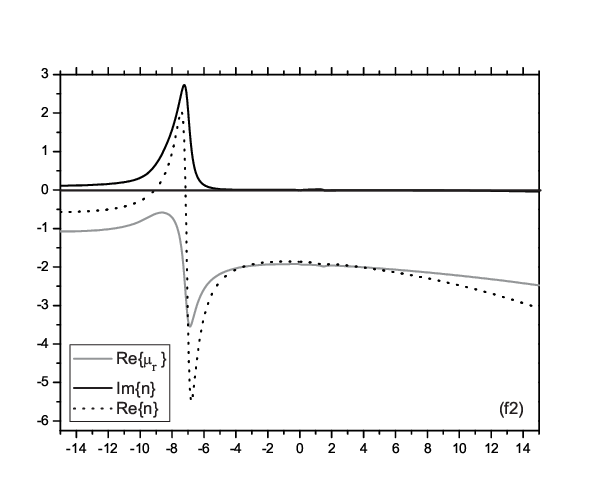}
  \hspace{0in}%
  \includegraphics[width=3.1in]{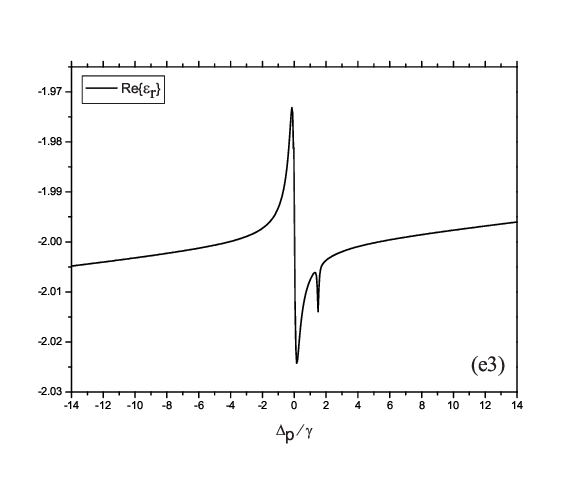}
  \hspace{0in}%
  \includegraphics[width=3.0in]{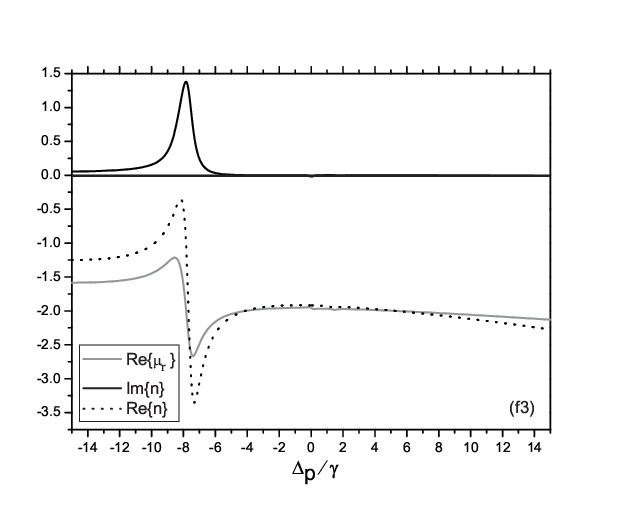}
  \hspace{0in}%
  \caption{Real parts of the permittivity and permeability,the refractive
  index as a function of $\Delta_{p}$/$\gamma$with: $\Omega_{p}$= $\Omega_{s}$=$0.1\gamma$, $1.0\gamma$,$2.8\gamma$,
  $\Omega_{c}$=$3.5\gamma$, $\Delta_{s}$=$1.5\gamma$. Other parameters are the same as those in Fig.2.}
\hspace{0in}%
\end{figure}\label{Fig.4}
\end{center}

In Figure 4, the signal field is set off-resonant
(e.g., $\Delta_{s}$=$1.5\gamma$) with the parameter
value:$\Omega_{c}$=$3.5\gamma$. We discuss the characteristics of
absorption and the refractive index by varying the the probe field's
Rabi frequencies ($\Omega_{p}$=$\Omega_{s}$) with 0.1$\gamma$,
1.0$\gamma$, 2.8$\gamma$.Other parameters are the same as those in
Figure 2. The profiles of $Re[\varepsilon_{r}]$ in
(e1)$\rightarrow$(e3) and $Re[\mu_{r}]$ in
(f1)$\rightarrow$(f3)display negative values. The simultaneous
negative permittivity and permeability indicates the left-handedness
of the atomic vapor medium. In (f1),the zero absorption intervals in
the solid curves are about [-3.6 $\gamma$, 0] and
[2.3$\gamma$, 5$\gamma$] when the $\Omega_{p}$ =0.1$\gamma$. The zero
absorption intervals are [-5.2$\gamma$, 14$\gamma$] and
[-5.5$\gamma$, 14$\gamma$] when the $\Omega_{p}$ =
1.0$\gamma$,2.8$\gamma$ in (f2) and (f3).The maximum of Re[n] in the
zero absorption intervals are about -3.2, -3, -2.4 in (f1)
$\rightarrow$(f3). Adjusting the intensity of probe beam properly,the
zero absorption characteristics of the left-handed material atomic
system can be manipulated with large flexibility.

\section{Conclusion}

In conclusion, we investigate three external fields interacting with
the four-level atomic system. Under the condition of coupling and
signal field being resonant simultaneously, the intensity of couple
field can expand the zero absorption interval but high intensity may
decrease the zero absorption ranges and negative refractive index of
the medium. When the two light fields are off-resonant, the intensity
of coupling field and the probe beam an also make no absorption in a
wider interval with a large negative refractive index. The zero
absorption property may be used to amplify the evanescent waves that
have been lost in the imaging by traditional lenses.Our scheme
proposes an approach to obtain negative refractive medium with zero
absorption and the possibility to enhance the imaging resolution in
realizing "superlenses". In our scheme, we can alter the value of the
refractive index while maintaining vanishing absorption to the beam.
In this sense, our approach implements modifying the optical response
of an atomic medium.

\section*{Acknowledgments}
The work is supported by the National Natural Science Foundation of
China (Grant No.60768001 and No.10464002).

\end{document}